# Combining sale records of landings and fishers knowledge for predicting métiers in a small-scale, multi-gear, multispecies fishery

Miquel Palmer[a,*], Borja Tolosa[a], Antoni Maria Grau[b], Maria del Mar Gil[a,c], Clara Obregón[a], Beatriz Morales-Nin[a]

[a] IMEDEA, Institut Mediterrani d'Estudis Avançats (CSIC-UIB), Miquel Marques 21, 07190 Esporles, Illes Balears, Spain

[b] Direcció General Pesca del Govern de les Illes Balears, Carrer Foners 10, 07006 Palma, Illes Balears, Spain

[c] LIMIA, Laboratori d'Investigacions Marines i Aqüicultura (Balearic Government), Eng. Gabriel Roca 69, 07157 Port d'Andratx, Illes Balears, Spain

* Corresponding author at:

Miquel Palmer

e-mail: palmer@imedea.uib-csic.es

Institut Mediterrani d'Estudis Avançats, CSIC/UIB

C/ Miquel Marques 21

07190 Esporles, Illes Balears, Spain

ABSTRACT

Stock management should be guided by assessment models that, among others, need to be fed by reliable data of catch and effort. However, precise data are difficult to obtain in heterogeneous fisheries. Specifically, small-scale, multi-gear, multispecies fisheries are dynamic systems where fishers may lively change fishing strategy (i.e., métier) conditioned by multiple drivers. Provided that some stocks can be shared by several métiers, a precise categorization of métiers should be the first step toward métier-specific estimates of catch and effort, which in turn would allow a better understanding of the system dynamics. Here we propose an approach for predicting the métier of any given fishing trip from its landing records. This approach combines the knowledge of expert fishers with the existing sales register of landings in Mallorca (Western Mediterranean). It successfully predicts métiers for all the 162,815 small-scale fishery fishing trips from Mallorca between 2004 and 2015. The largest effort is invested in the métiers Cuttlefish/Fish and Spiny lobster, landings peak for Cuttlefish/Fish and Dolphinfish and revenues for Spiny lobster and Dolphinfish. Métier predictions also allowed us to describe the temporal (seasonal and between-year) trends experienced by each métier and to characterize the species (commercial categories) that are specific to each métier. Seasonal variability is by far more relevant than between-year variability, which confirms that at least some fishers are adopting a rotation cycle of métiers along the year. Effort (fishing trips), landings and gross revenues decreased in the last 12 years (2004 to 2015). The approach proposed is also applicable to any other fishery for which the métier for a fishing trip sample is known (e.g., on-board observers or logbooks), but relying on fishers expertise points more directly to fishers’ intention. Thus, métier predictions produced with the proposed approach are closer to the actual uses of fishers, providing better grounds for an improved management.

## 1. Introduction

Most fisheries are not homogeneous but consist of a variety of vessels and activities that differ greatly in terms of, among many other factors, vessel size, gears used, technology employed, fishing grounds reached, and degree of expertise of the fishers. All these factors are also highly dependent on the market characteristics the fishery delivers to, and on a range of social aspects such as local culture and the availability of capital investment (Therkildsen, 2007). While all these factors are affected by the targeted fish stocks, they are also affecting the stocks themselves.

Conventional fisheries data collection, advice, and management usually target single-stocks. At this basis, assessing fishing mortality throughout the relationship between catch and effort may be affordable for homogeneous, monospecific fleets. Nevertheless, this approach has long been recognized as inadequate when applied to heterogeneous fisheries, which are subjected to interactions between subsets of fishing units (e.g., métiers), and across species (Marchal, 2008). Biased estimation of fishing mortality may result from naïvely pooling catch/effort across heterogeneous units (i.e., ignoring between-métier differences). This fact is recognized, for example, for multi-fleet, multi-species bio-economic models (e.g., MEFISTO model; Lleonart et al., 2003), which are specifically designed for including specific input of effort and catchability for every fishing unit (e.g., métier) considered. Accordingly, not only more accurate predictions of the stock dynamics can be obtained but also better predictions for different métier-specific management decisions can be provided.

Several steps have been undertaken in the past to explicitly incorporate the heterogeneity of the fishing activities within the cycle of observing, assessing, forecasting, and managing fisheries. A common sense solution is to identify units as homogeneous as possible. ICES (the International Council for the Exploration of the Sea) considers three types of fishing unit: the fleet, the fishery, and the métier (ICES, 2003). A fleet is a group of vessels sharing similar characteristics in terms of technical features and main activity. A fishery is a group of fishing trips targeting the same assemblage of species/stocks, using similar gear, during the same period of the year and within the same area. Nevertheless, fleet and fishery are often too heterogeneous from a managing perspective. Conversely, the concept of métier is specifically aimed to define a homogeneous subdivision of, either a fishery by vessel type or a fleet by type of fishing trip. Specifically, a métier is characterized by the use of a single gear targeting a specific group of species. It usually operates in a given area during a given season, within which each boat deploys a similar exploitation pattern; i.e., the species composition and size distribution of the catches taken by any vessel working in a particular métier will be approximately the same (Alarcón-Urbistondo, 2002; Deporte et al., 2012; Mesnil and Shepherd, 1990). Provided that different métiers can share several target stocks, the total effort and catches upon a stock can only be properly estimated after combining all the involved métiers targeting this stock.

In the Mediterranean, the small-scale fleet (SSF) is very relevant socially, economically and has a long history (Stergiou et al., 2006). Around 80% of the registered boats in the European Mediterranean belong to this fleet and these are currently employing about 100,000 people (Maynou et al., 2011). The number of small-scale boats operating at the Spanish Mediterranean has been estimated in 1,462 in 2015 (STECF, 2016). The fleet in the Balearic Islands (GSA05; Geographic Sub Area 5 of the General Fisheries Commission for the Mediterranean) follows a similar pattern: it comprises 337 boats, being 85% small-scale (which employ 700 people), 12.5% trawlers and the rest corresponding to different modalities (data for 2012; Grau et al., 2015). It is noticeable that the number of small-scale boats is experiencing a decreasing trend (345 boats in 2009 and 278 in 2015; Grau et al., 2015). This trend may result from both, the implementation of measures aimed to reduce effort and the decrease of fish price (Morales-Nin et al., 2010). Nevertheless, this trend is impaired with landings, which remain around 400 tons/year (Morales-Nin et al., 2010).

A peculiarity of most SSF is that some boats may use several fishing systems, which are lively alternated during the year according to the availability of resources, market demand, and other factors, such as management policies (e.g., closing seasons), local environmental characteristics and interaction with other fishing gears (Maynou et al., 2011; Salas and Gaertner, 2004). Therefore, SSF not only constitutes a relevant fraction of the fishing activity in some areas but also is particularly heterogeneous and thus, challenging from a managing perspective.

Despite its importance worldwide, SSF practices have been generally subject to little attention by the scientific community and managers when compared to the industrial fishing sector. Therefore, there is an objective need for delineating métiers in such fisheries. However, this is in practice a more challenging goal than expected. The approaches used in the past to identify métiers either (i) make use of existing records on the technical features of fishing trips (e.g., gear and mesh size used, fishing grounds visited, season, boat characteristics), which may be available from fishers' logbooks, (Marchal et al., 2006; Ulrich et al., 2001), or inferred from interviews with fishers (Christensen and Raakjær, 2006; Neis et al., 1999), or (ii) are intended to ascertain the métiers used by retrospectively examining the landings (Deporte et al., 2012; Marchal, 2008).

In this paper, we propose to combine some of the advantages of all these approaches. Using the small-scale fleet from Mallorca Island (Western Mediterranean) as case study, we demonstrated how fishers' expertise can be combined with the relatively recent implementation of electronic register of landings in order to elucidate the métiers practiced by a particularly heterogeneous fleet. The specific aim of this work is not only to select and test a numerical algorithm for predicting the métier a given boat has practiced from the corresponding sale record, but also to up-scale the predictions of métiers to the entire fleet, which will provide an accurate, quantitative description of catches and effort for each métier. Thereby, more precise assessment of the fishing mortality of all the exploited stocks by the small-scale fishery will be obtained after more precise delineation of the métiers, which in turn should contribute to improve the management of the fishery. Information on the gear/fishing tactic, the main species exploited, the characteristics of the vessels involved plus background on the métier-specific temporal trends in catch, effort, gross revenues, as well as between-métier interactions for the period 2004-2015 are provided.

## 2. Material and methods

### *2.1. Métiers and data*

The small-scale fleet in Mallorca is conducted by vessels less than 10 gross tons (Decree 17/2003 of February 21, from the Balearic Islands Government), with 1-3 hand decks, and operating close to the base harbor. This definition is consistent with other EU level definitions such as the Council Regulation (EC) No 1198/2006 of 27 July 2006 ('small-scale coastal fishing' means fishing carried out by fishing vessels of an overall length of less than 12 meters and not using towed gear as listed in Annex I of Commission Regulation (EC) No 26/2004 of 30 December 2003 regarding the fishing vessels register of the Community). Specifically, trawlers and large seiners are not considered here as SSF. Less than 1 day outings are compulsory and some combinations of fishing gears in the same fishing trip are not permitted. More details on the fishery management on this fleet are provided elsewhere (Morales-Nin et al., 2010).

In Mallorca, the commercialization of all the landings (i.e., SSF, trawlers and seiners) is made through a single central fishing wharf (*OPMallorcaMar*), which is a cooperative composed by all the boat's owners in the island. In addition, fishers are associated in guilds by port (*Confraries*), which in turn are associated in the Balearic Fishers Federation (*Federació Balear de Confraries de Pescadors*). The landings are arranged in standard boxes by the fishers and auctioned daily in decreasing prices. An automatic selling procedure, implemented since 2004, registers for each box, among other data, the commercial category, the weight in kilos, the price and the name of the boat. The personal data of the fisher are encrypted in accordance with the terms of a confidentiality agreement.

The time series (2004-2015) resulting from the daily sale records, provides a valuable information on the fishing activities, and how they change at different time scales (e.g., seasonal and decadal). However, some potentially confounding factors hinder the usefulness of such database for fishing management. For example, some species might be sold as more than one commercial category, e.g., small, medium and large hake (*Merluccius merluccius*); boxes with mixed catches can correspond to different commercial categories depending on the vim of the auctioneer; and boats could have changed their name (and owner) along the time series of data. Nevertheless, one of the major drawbacks is that the métier used for obtaining the catch is not provided.

We propose to use fishers' knowledge in order to infer the métier for any fishing trip. The proposed strategy (Fig. 1) started with selecting a representative sample of fishing trips. The list of catches (i.e., the list of the daily sales of a given boat) of those sampled fishing trips were then presented to a number of experts (fishers), who were asked to label them with the métier/s (from a closed list) that they thought had most probably been used to get such a combination of catches. This sample of labeled records was then used for selecting, parametrizing and testing the success of a range of classification algorithms. Finally, the best algorithm was used for up-scaling métier's predictions from the sample to the entire time series (2004-2015) of sales from the daily boat records.

The specific details of the full data mining and analysis are summarized in Fig. 1. The steps were:

1) The information for a given day was received as an ACCES archive that is structured by fish box (each row contents the data from a single box). These files were automatically read using the *RODBC* library (Ripley and Lapsley, 2015) from R (https://www.r-project.org/) and stored.

2) In general, several of the fish boxes above correspond to the same boat and they must be restructured to summarize the fishing activity of one boat in a day (i.e., one fishing trip). Therefore, the box data were restructured into a matrix composed by rows consisting in all the fish sold by a boat in a day, and the commercial categories in columns. Two separate matrices were produced for fish weight (kg per fishing trip) and gross revenues (Euros per fishing trip). Provided that fishing trip duration is one-day maximum, and that all the fish landed in a given day is auctioned off after a few hours, any row in the restructured matrix corresponds to all the catches landed by a boat in a single fishing trip (day). The data were cleaned of negative sales, which represent errors in the purchase or devolutions.

3) The matrix of daily boat records above included trawlers and large seiners that must be filtered out. Provided that boat category (namely, small-scale, trawler or large seiner) was available for all the active boats in 2012, an auxiliary classification algorithm was implemented using a random forest, as implemented in the *random Forest* library (Liaw and Wiener, 2002) of the R package. The performance of such algorithm was tested by cross validation and used to predict the boat category of all the fleet (2004 to 2015).

4) The results of the filtering step above were two matrices (either for weight or gross revenues) of rows consisting in the daily boat's sales records for the SSF only (162,815 rows; see Section 3.1), and covering from 2004 to 2015. The columns were the 75 (see Section 3.1) commercial categories considered after removing very uncommon categories.

5) A total of 1,550 daily boat records were sampled from the weight matrix in Step 4. The weight matrix was preferred to the gross revenues matrix because seasonal trends of average fish price would confound landing trends. Provided that métier prevalence seems to be largely unbalanced, an *ad-hoc* sampling strategy was adopted to avoid underrepresentation of the less frequent métiers. The weight matrix was submitted to a principal component analysis, the first 10 axes were divided into 10 segments of the same length and, finally, a number of samples proportional to the variance explained by each of those 10 axes were then randomly selected from each of the corresponding segments.

6) Independently of the landing data (Steps 1 to 5), a preliminary list of the 12 main métiers currently used in Mallorca (Table 1) was drawn up after combining bibliographic (Massuti, 1995), legal normative (Regulation of Artisanal Fisheries in Balearic inland waters, Decree 17/2003, Balearic Official Bulletin 28, 01/03/2003) and face-to-face data, by carrying out unstructured interviews to five very experienced fishers.

7) A new panel of 15 expert fishers was then selected. Each one of these 15 experts was asked to label a random (see Step 8 below) subsample of 150 daily boat records selected from the total records described in Step 5. Each daily record must be labeled with one or more of the 12 métiers described in Step 6 (i.e., from a closed list). The information available to the fishers was the species list within the catch, the weight per species and the date. The questionaire was usually completed whithin around 30-45 minutes.

8) To test the coherence between each expert when labeling daily boat records with métier/s, 50 of the 150 records in Step 7 were the same for all experts. Provided the nature of the response matrix (0/1 of 12 métiers), a canonical correspondence analysis (CCA), implemented using the *cca* function of the *vegan* library (Oksanen et al., 2014) from R, allowed to test between-fishers differences (Borcard et al., 2011). The initial 15 experts were submitted to an expert-by-expert sequential elimination protocol that continued until the remaining experts' set showed non-significant differences between them. Differences were tested using bootstrapping. Specifically, experts were randomly permuted while the order of the daily boat records was kept fixed (i.e., constrained bootstrapping as implemented in the *vegan* library).

The remaining data set after removing non-consistent experts were used to select and test a classification algorithm. Full details of the classification strategy are provided in the Section 2.2.

Quality control of the prediction success of the classification algorithms was assessed by cross validation. Specifically, some of the initial 12 métiers were found either uncommon or experts were unable to successfully discriminate between some pairs of similar métiers. The outcome was that success classification rate of these métiers were low. Therefore, some of the initial 12 métiers were collapsed or deleted, which results in a final list of métiers. The resulting data set (i.e., coherent experts and finally retained métiers) were used for parameterizing and re-testing the classification algorithm finally selected.

9) This final algorithm was used to up-scale métier's predictions to the full time series of daily boat records covering from 2004 to 2015, which allowed us to describe the seasonal trends of effort (fishing trips), landings (kg) and gross revenues (€). The commercial categories of landings that best characterize the métiers finally considered was assessed using SIMPER analysis (Clarke, 1993), as implemented in the *simper* function of the *vegan* library from R.

### *2.2. Predicting métier from the sales daily boat record*

Classification problems in which an object can be classified into more than one category are known as multi-label classification problems (Tsoumakas and Katakis, 2007). Conventional classification methods, either parametric (e.g., discriminant analysis) or non-parametric (e.g., *random forest*) assume that categories are mutually exclusive (Jones et al., 2017). Two approaches for adapting the conventional methods to multilabel classification have been proposed. *Label powerset* defines a new (mixed) category for the cases a fisher uses two métiers during the same fishing trip. *Binary relevance* consists, for a given métier, in splitting the data into two categories: the daily boat records assigned versus non-assigned to the métier considered. Therefore, the problem is transformed in a number (as many as métiers) of binary classification problems.

Conventional classification methods can be applied in both cases after adapting the input data. After preliminary inspection of the data structure, *binary relevance* was selected because the success of *label powerset* depends on the assumption that any mixed categories must be well represented, which is not the case in our data (see Results, Section 3). Preliminary inspection and data recoding were completed using the *mldr* library (Charte and Charte, 2015) from the R package.

Recoded data were evaluated using all the suitable methods implemented in WEKA (http://www.cs.waikato.ac.nz/ml/weka/). WEKA (Witten and Frank, 2005) implements an exhaustive collection of machine learning algorithms within which there are 17 binary classification algorithms. The algorithm selected was which showed the largest cross-validated predictive capability as measured using kappa index (Jones et al., 2017). This process was fully automatized using the *Rweka* library (Hornik et al., 2009) from R.

### *2.3. Comparison with the conventional method*

The conventional, alternative approach is aimed to infer the métiers used by retrospectively examining the catch profiles resulting from fishing trips (Deporte et al., 2012; Marchal, 2008). The raw data (or the transformed data after applying some multivariate analysis for reducing dimensionality) is clustered and a cut-off level that minimizes within cluster heterogeneity is selected. Here, *clara* clustering algorithm (as implemented in the cluster library from R; Maechler et al., 2015) was applied to a random sample of 10,000 daily boat records. The optimal number of clusters was selected using the *silhouette* function of the cluster library, after comparing the results obtained with 2 to 15 groups (i.e., métiers). This function computes the average cohesion (how similar an object is to its own cluster in comparison with other clusters), thus it provides the metrics for assessing if there are too many or too few clusters.

## 3. Results

### *3.1. Cleaning the input data matrix*

The huge raw dataset, consisting in more than 5 million fish box entries (Step 1; Fig. 1), was successfully reorganized in two matrices (weight and gross revenues) (see Step 2; Fig. 1). The 256,490 rows of these matrices were the daily boat records from 2004 to 2015 of all the fleets (i.e., including small-scale, trawlers and big seiners). The columns were all the occurring commercial categories (170).

A random forest classification tool for predicting broad categories (namely, small-scale, trawlers and seiners) was successfully implemented with the 2012 subset of data (see Step 3; Fig. 1). As cross-validated success of predictions was excellent (Table 2; kappa index = 0.99), this tool was used to predict the broad category for the full time series (2004-2015). The number of small-scale daily boat records for the 2004-2015 (Step 4; Fig. 1) was 162,815 (or 63%). After removing trawl- and seine-specific commercial categories, plus a few commercial categories with very low prevalence for SSF, a total of 75 commercial categories were retained.

A total of 1,550 daily boat records were selected from SSF (see Step 5; Fig. 1), and these were labeled by fifteen expert fishers into the twelve métiers listed in Table 1 (see Step 7; Fig. 1). When experts were found to differ, multivariate (CCA) scores were plotted and the expert showing the most disparate pattern of métier predictions was deleted. As an example, the initial step (15 experts: $F = 2.04$, $df = 14{,}735$, Prob. <

0.001) is shown in Fig. 2. This loop was repeated until no between-expert differences were found, which allowed to successfully detect four outlier experts (i.e., 11 out of 15 experts were retained; F = 0.34, df = 10,539, Prob. = 0.111).

After removing those experts, a reliable métier prediction is assumed for the remaining 1,100 daily boat records. Nevertheless, the cross-validated classification success of up to 17 algorithms leaved room for improvement in five out of the twelve métiers initially considered (Step 6; Fig. 1). In most of the cases, these métiers are old-fashioned and underrepresented, thus machine learning algorithms seem unable to build a reliable model. Therefore, they were merged or deleted. The case of the métiers Cuttlefish (targeting *Sepia officinalis*) and Fish (mixed fish) deserves particular attention because there are two trammel nets that differ by legal normative in mesh size and in seasonality (Regulation of Artisanal Fisheries in Balearic inland waters, Decree 17/2003, Balearic Official Bulletin 28, 01/03/2003). Yet, expert fishers were not able to efficiently discriminate between them, suggesting that catches may be quite similar, at least during a certain period of the year. Therefore, these two métiers have been merged in a single category (Cuttlefish/Fish).

### *3.2. Classification algorithm*

The cleaned data set consisted in 1,100 daily boat records, for which a reliable métier prediction is available among a closed list of seven possible métiers (Cuttlefish/Fish, Transparent goby, Dolphinfish, Squid jigging, Spiny lobster, Red mullet and Longline; Table 1), were re-tested using the same seventeen WEKA classification algorithms. The algorithm showing the best performance (after cross-validation) was *IBk*. This algorithm implements a k-nearest neighbor classifier (Fix and Hodges, 1951). Membership prediction of a new object depends on the membership of the majority of their k-nearest neighbors. The confusion matrices for each one of the 7 métiers considered are shown in Table 3. In four cases (Transparent goby, Dolphinfish, Squid jigging and Longline), the percentage of correct predictions was perfect (100%) and for the remaining three métiers the number of failures was negligible.

Provided that the cross-validated predictions of the classification algorithm were excellent, this method was used for predicting the métier that most plausibly had been used in each one of the 162,815 fishing trips carried out for the Majorcan SSF between 2004 and 2015 (Step 9; Fig. 1). The relative importance of the 7 main métiers both in effort, landings and gross revenues are summarized in Table 4. Note that the total effort may be higher than the number of daily boat records due to the simultaneous use of more than one métier per day by some boats. The largest effort was invested in Cuttlefish/Fish (mean ± sd; 4,964.3 ± 600.1 fishing trips/year; n = 12 years). The main contributors to the landings were Cuttlefish/Fish (122.9 ± 19.1 tons/year; n = 12 years) and Dolphinfish (109.4 ± 33.7 tons/year; n = 12 years), albeit the main income corresponds to Spiny lobster with a mean value of 0.99 ± 0.1 million euros/year (from now on MEuros; n = 12 years) due to its higher price, followed by Cuttlefish/Fish (0.96 ± 0.1 MEuros/year; n = 12 years) and Longline (0.7 ± 0.1 MEuros/year; n = 12 years). Note also the disproportionately lesser effort invested in Dolphinfish (8.2%) in front of Cuttlefish/Fish (33.9%), which is related with the seasonal pattern of resource exploitation (Section 3.3).

*3.3. Temporal and seasonal trends*

The temporal patterns displayed for the SSF along with all the considered period (2004-2015) are displayed in Fig. 3. Based on the values estimated after fitting the annual data (12 years; 2004 to 2015) to a linear trend, we can conclude that landings of the entire SSF have decreased in 23%, from 526 tons in 2004 (95% confidence interval, CI: 407 to 645) to 401 tons in 2015 (282 to 520). The monthly trend for landings is displayed in Fig. 3a. The decreasing trend for effort (number of fishing trips of the entire SSF) was 20%, going from 16,280 fishing trips in 2004 (95% CI: 14,708 to 17,850) to 13,044 fishing trips in 2015 (11,473 to 14,616). The monthly trend for effort is displayed in Fig. 3b. Finally, the decreasing of gross revenues in nominal terms (i.e., without adjusting for inflation) was 18%, from 4.5 MEuros in 2004 (95% CI: 4.2 to 4.9) to 3.7 MEuros in 2015 (3.3 to 4.1). Note however, that the actual decrease in real terms (i.e., after adjusting for inflation) was 33%. The monthly trend for gross revenues is displayed in Fig. 3c. Contrasting with those decreasing trends, CPUE (landings per fishing trip; Fig. 3d) seems to be stationary (the slope of a linear trend of annual averaged CPUE was not significantly different from zero; Prob. = 0.57; mean ± sd; 31.5 ± 2.3 kg per fishing trip; n = 12 years). Gross revenues per fishing trip in nominal terms (Fig. 3e) seem stationary too (slope of a linear trend of annual averaged data was not significantly different from zero; Prob. = 0.59; mean ± sd; 283 ± 12 euros per fishing trip; n = 12 years).

Métier-specific patterns for the period 2004-2015 are depicted in Fig. 4. Overall, the fishers strategies (i.e., relative importance of the different métiers at the year scale) have not experienced large changes over time. Between-year variations (e.g., high of the seasonal maximums) are small, excepting in the case of the landings of Dolphinfish. Contrasting with such a relatively small variability at the decadal scale, seasonal variability is very important (Fig. 4). The periodicity at the métier level was precisely regular across years, for effort, landings and gross revenues. Concerning effort, when the predictions were polled by month across the 12 years considered (Fig. 5), the resulting pattern supports the hypothesis that some métiers are seasonally rotated. The canonical cycle was already described (Iglesias et al., 1994) but here a more precise delineation of the métier-specific periodicity is provided. The cycle starts with Transparent goby in winter, followed by Cuttlefish/Fish (but see below) in spring, Spiny lobster in summer and Dolphinfish in autumn. The other métiers did not show a so clear seasonal pattern or were carried out with similar intensity along the year (e.g., Longline). The extended exploitation pattern predicted for Cuttlefish/Fish from August to December was due to the inability of the experts to discriminate between Cuttlefish and Fish (Fig. 5).

Finally, the results of the SIMPER analysis completed for identifying the commercial categories that better characterize (in terms of landed weight by category) each one of the 7 métiers is detailed in Table 5. In three cases (Transparent goby, Dolphinfish and Squid jigging), landings are composed by the target species (*Aphia minuta*, *Coryphaena hippurus* and *Loligo vulgaris* respectively), plus a very few secondary categories of commercialized by-catch. The case of Squid jigging is noticeable because this métier may be a secondary activity: fishers would target squid (*Loligo vulgaris*) while using another static métier that forces them to wait for the catch. In this case, the squid gear used (hand line) is very selective and renders few but high quality and high valued product. The same squid species reaches a lower price when captured with other gears (i.e., trawling).

Conversely, Cuttlefish/Fish, Spiny lobster and Red mullet are set nets (trammel nets and gillnets) characterized by a long list of by-catch in addition to the main target species, which are *Sepia officinalis*, *Palinurus elephas* and *Mullus surmuletus*, respectively. However, such a by-catch is not only a relevant fraction of the landings but also of the gross revenues, especially in the case of Spiny lobster. Finally, the landings of Longline (also called *Palangró*) are largely unspecific, being large sparids and serranids (e.g., *Epinephelus marginatus*) the most valued of the target species.

### *3.4. Comparison with the conventional method*

The conventional method showed that the optimal number of groups (i.e., métiers) in which a random sample of 10,000 daily boat records can be optimally split after clustering the landings profiles is only two. The silhouette index peaked at two groups and showed an irregular but decreasing trend while increasing the number of potential métiers (see Fig. 6). After analyzing the species composition of these two clusters, the first one seems to fit well with the Dolphinfish métier, but the second is a mix of the other six métiers considered here. Therefore, in our case it seems that métiers cannot be unequivocally assessed by clustering the landings profiles, irrespective of the technicalities applied (e.g., with or without applying a preliminary multivariate analysis for reducing dimensionality, the distance/dissimilarity metrics or the clustering algorithm).

## **4. Discussion**

### *4.1. Toward a pragmatic métier concept*

The results reported here support that fishers are able to successfully classify fishing trips into discrete units using landings only. We propose that it is possible because landings reflect fisher's intention (e.g., métier choice or even specific fishing tactics), which can be accurately inferred from fisher perception (i.e., the fisher believe on another fisher intention given only the landings of a trip). We propose that such ability is the result of fisher's experience.

The decreasing trend on the number of boats and gross revenues in the last decades, and the small average gross revenues per trip reported here supports that most small-scale fleets may be close to economic sustainability. Accordingly, fishers have to be continuously adapting fishing strategies and tactics. Thus, impelled by market demands, experienced fishers (i.e., those that remain in the activity because they successfully compete) have learned how to modify fishing strategy (e.g., métier choice) and tactics (any specific detail of the fishing strategy) for maximizing the likelihood of obtaining the desired landings (i.e., those that achieve better price). In the same way but opposite direction, experienced fishers are able to consistently cluster fishing trips into métiers based on landings only, even when the signal provided by landings is very noisy.

Provided that fisher intention (e.g., métier choice) can be accurately inferred from fisher perception based on landings only, all fishing trips that are classified in a given group are sharing very close fishing strategies and tactics. Therefore, the natural units for structuring management decisions (i.e., the métiers) should be the units that fishers can consistently recognize. Stock assessment based in units that accurately and precisely reflect the actual uses (i.e., fisher intention) of the fleet would better predict any threat for sustainability, thus allowing the implementation of precise (i.e., métier-specific) management rules.

Therefore, the framework proposed here allows defining métiers in a more pragmatic way, and describing them comprehensibly for management purposes. We suggest a new method to predict the métier for any fishing trip from historical time series of landings data. Briefly, métier prediction for the entire SSF is made possible after training a classifier with a sample of landing records that has been classified into métiers by a panel of expert fishers (Fig. 7). This strategy points more directly to fishers intention and, to our knowledge, this is the first time that experts' knowledge is combined with a sale record of landings in such a way. An obvious limitation of the method is that a landing record for each fishing trip should exist already. Nevertheless, its implementation is promoted by the EU (Marchal, 2008), hence the proposed framework might expand within other countries.

The conventional approach for inferring métiers from landing records has been to cluster fishing trips according to their similarity in landing composition. The rationale behind is that the clusters obtained include the fishing trips in which the same métier was used (Alemany and Álvarez, 2003; García-Rodríguez, 2003; Tzanatos et al., 2006). This approach has two practical drawbacks (Palmer et al., 2009): (i) to objectively determine the optimal number of clusters and (ii) to unequivocally define a métier for a given cluster, which is carried out *a posteriori*. That is, the most common species in the landings profiles from a cluster are assumed to characterize the (single) métier used for all the fishing trips in that cluster. In addition, the unit of the landing records should be the fishing trip because when pooling units together (Alemany and Álvarez, 2003; García-Rodríguez, 2003; Poulard and Leaute, 2002), any subjacent variability will be confounded.

In the case of SSF from Mallorca, the conventional clustering approach seems unsuccessful because only two métiers can be objectively identified (Fig. 6). The large variability of catches (e.g., up to 75 commercial categories and a striking seasonal pattern) is the possible cause behind such a failure. Even thought that the clustering approach has represented a relevant improvement and it is being successfully used in some fisheries (Deporte et al., 2012; Marchal, 2008), it may be suboptimal for heterogeneous fisheries, as most SSF are.

Palmer et al. (2009) proposed the use of a sample of on-board observations for training conventional and machine-learning classifiers with landings profiles. The framework suggested here (Fig. 7) goes a step further, in that on-board observations (i.e., the observer selects a métier from an *a priori* defined and closed list of métiers) has been changed by the perception of expert fishers on what métier has probably been used for obtaining a given landings profile.

In summary, accordingly with the key role of fishers intention, we propose a more pragmatic métier delineation as the unit that fishers can consistently recognize. Similarly, the optimal number of métiers in which a fleet could be divided and should be managed will be those that fishers can consistently recognize. The outcomes of this paradigm change are discussed in the two next sections.

### *4.2. Practical advantages, limitations and technicalities*

The method proposed here (Fig. 1) is able to accurately predict the métier that most plausibly has been used in a given fishing trip. The excellent cross validated prediction success (close to 100%) for a sample of 1,100 boat daily records reinforces the reliability of up-scaling predictions from such a sample to the entire SSF and for the period from 2004 to 2015.

Concerning the technicalities of the classification protocol, it is noteworthy that some of the expert fishers which labeled boat daily records may behave in a different way, thus the need of a strict quality control is fully justified. The sequential removal of outlier experts used here is a straightforward option as it ensures between-expert coherence while minimizing the number of excluded experts. Another relevant technicality is the use of an adequate data-mining platform. In this regard, the R libraries have provided an invaluable support because any step of the data-mining process can be easily structured in a single ad-hoc script. Finally, the use of classification methods that allow for multiple labeling of the same object (Tsoumakas and Katakis, 2007) has been decisive as in most SSF more than one métier can be used during a single fishing trip. Nevertheless, in the case of the Balearic Islands, only some combinations are allowed. Therefore, multiple assignments can reflect either uncertainty of the expert fishers when assigning métier or the fact that more than one métier have been actually used. Specific improvements on discriminating those cases should be desirable.

Replacing on-board observers by a panel of expert fishers represents an additional advantage: extensive programs of on-board observations are relatively common in trawlers and other large vessels but sporadic in SSF. Moreover, EU vessels under 10 m are not required to report any logbook. Interview surveys have been used to collect quantitative information (Marchal, 2008) but provided the large number of participants in SSF, to engage a representative panel of expert fishers may be a better and economically more affordable option.

The use of landing profiles for defining métiers has been criticized because it does not consider discarding (Marchal, 2008). Certainly, sale records comprise, in addition to the main targets, the by-catch fraction that is commercialized only. However, no information is available on the relevance of the discards or on their composition by métier. On-board observations are thus unavoidable to quantify discards. In the case of the SSF from Mallorca, a specific on-board sampling program is in progress (Program EU Horizon 2020 Research and Innovation Action SFS-634495). Therefore, the results of these on-board samples of discards will allow us to up-scale the discards per métier to the entire fleet after taking into account the effort per métier provided here.

In the specific case of the Mallorcan SSF, expert fishers were not able to efficiently discriminate from the reported catches which of two specific trammel nets (Cuttlefish and Fish) had been used. These nets actually differ by legal normative in mesh size and in seasonality (Section 3.1) but the fishing grounds are similar. Cuttlefish nets are usually set at the lower boundary of the seagrass *Posidonia oceanica* and Fish is not as habitat-specific. Catch composition may smoothly change in a way that fishers are not able to discriminate between them. In those cases, additional information of the size distribution of the catches will be decisive for properly splitting a currently merged métier (Cuttlefish/Fish). To solve this situation, the managing authorities (*Direcció General de Pesca del Govern de les Iles Balears*) and the IMEDEA are launching a specific monitoring program.

### *4.3. Management outcomes*

Small-scale fisheries represent more than half of the world's annual marine fish catch of 98 million tonnes (Berkes et al., 2001), contributing with 0.3 million tonnes in EU (STECF, 2016). However, they are usually left behind industrial fisheries, mostly due to the lack of data regarding stock trends or the fishery's socio-economic impacts (Stergiou et al., 2006). This partly explains some of the constraints faced by SSF (FAO, 2005-2015). In the Mediterranean, SSF performs a relevant fraction of the effort. The data here reported suggests that 63% of the fishing trips are operated by SSF. Therefore, albeit trawlers and seiners provided most of the landed fish in Mallorca (Quetglas et al., 2016), the direct and indirect economic impact of SSF are noticeable (Carreras et al., 2015). According to Quetglas et al. (2016), SSF represents 20% of landings and 27% of gross revenues. Moreover, the quality of the product provided (and consequently the gross revenues obtained) by SSF in Mallorca is excellent (Morales-Nin et al., 2013). Hence, this activity has an important socio-economical relevance (Maynou et al., 2013; Morales-Nin et al., 2010).

Nonetheless, the number of small-scale boats in Mallorca has been experiencing a relevant decrease in the last decades (Section 1). The data reported here suggests that effort, landings and gross revenues for the entire small-scale fleet are decreasing too (Section 3), which proves the weakness of the system and suggests that it may collapse in the near future if not properly managed. The mean age of the fishers engaged and the low replacement rate points at this direction (Maynou et al., 2013). Thus, in this case, an eventual fisheries collapse may not be directly related with resource status. The trends at a boat level cannot be properly analyzed due to a confidentiality agreement, yet it will be very interesting to disentangle if the less efficient boats are those that are quitting the activity (fishers sorting). In this case, apparent stability of CPUE (landings per fishing trip) may mask a decrease in stock abundance (van Poorten et al., 2016). Alternatively, warning signs should not be deducted from stationary CPUE.

In contrast with such difficult perspectives, further efficient management measures as co-management are already being implemented. This approach may be particularly advantageous for small-scale fisheries. In the Mediterranean, this strategy has been suggested as promising since fishers seem prone to adopt it (Lleonart et al., 2014). For example, in the case of the spiny lobster fishery from Mallorca, fishers would agree in maximizing profits instead of catches (Amengual-Ramis et al., 2016).

The case of the Mediterranean sand eel (*Gymnammodytes cicerellus*) fishery is a positive example of a new way of managing a resource because it allows the fishers to control their own fishery, with the help of scientists (marine biologists and socioeconomists), policy makers and NGOs (Lleonart et al., 2014). A similar approach has been implemented in Balearic Islands for the transparent goby fishery (Morales-Nin et al., 2017).

In this scenario, the concept of métier and the analytical approach provided here became even more relevant. The fact that métiers are defined and delimited by the local fishers suggests an, only apparent, drawback: the set of defined métiers and their limits (i.e., their characteristics in terms of landing composition) will be useful only at the local scale. Nevertheless, spatially well delimited stocks as those exploited by SSF in Mallorca (the Balearic Islands waters being encompassed within a single GFCM area, GSA05) will be better managed at a local scale (Quetglas et al., 2012). Obviously, this is not the same case of transnational stocks, exploited by several, well differentiated fleets, for which a transnational métier concept as the one suggested by Deporte et al. (2012) would be a better option.

In summary, we suggest a local management of the SSF from Mallorca. The new métier definition proposed and the analytical tools provided here are more appropriate for this management scale. The precise categorization of métiers given here should be the first step towards métier-specific estimates of catches, effort and gross revenues, which in turn, should allow a better understanding of the drivers of system's dynamic (i.e., the drivers of fisher decisions on the specific métier to be adopted at any moment). In this scenario, co-management would receive better scientific advice and would have more chance for success.

## Acknowledgements

This is a result of the Associated Unit LIMIA-IMEDEA. This work has received funding from the European Union's Horizon 2020 Research and Innovation programme under Grant Agreements 634495 (MINOUW) and 678193 (CERES). *OPMallorcaMar* and *Federació Balear de Confraries de Pescadors* are thanked for their support and collaboration. We specially thank all the expert fishers who answered the questionnaires. Eugenio Garcia helps us during the first stages of this work. Francesc Riera and Francesc Maynou contributed to discussion.

| MÉTIER (Local name) | Gear group | Target assemblage | Mesh size/gear characteristics | Target species | Activity period |
|---|---|---|---|---|---|
| **Cuttlefish/Fish (*Sipia/Peix*)** | **Bottom trammel net** | **Benthic assemblages** | **67 mm/ max. 4.500 m per vessel** | **Cuttlefish (*Sepia officinalis*), mixed fishes** | **Winter** |
| **Spiny lobster (*Llagosta*)** | **Bottom trammel net** | **Benthic assemblages** | **130 mm/ max. 4.500 m per vessel** | **Spiny lobster (*Palinurus elephas*)** | **Spring-Summer** |
| **Red mullet (*Moll*)** | **Bottom trammel net** | **Benthic assemblages** | **50 mm/ max. 4.500 m per vessel** | **Red mullet (*Mullus surmuletus*)** | **Summer - Autumn** |
| **Longline (*Palangró*)** | **Bottom long-lines** | **Benthic assemblages** | **Hooks min. 9 mm wide /max. 1.000 hooks per vessel** | **Red porgy (*Pagrus pagrus*), red scorpionfish (*Scorpaena scrofa*)** | **All year** |
| **Transparent goby (*Jonquillera*)** | **Special pelagic surrounding net** | **Pelagic fishes** | **- / max. 100 m** | **Transparent goby (*Aphia minuta*)** | **Winter** |
| **Dolphinfish (*Llampuguera*)** | **FADS and special surrounding net** | **Epipelagic fishes** | **- / max. 200 m** | **Dolphinfish (*Coryphaena hippurus*)** | **Autumn** |
| **Squid jigging (*Potera*)** | **Hand line** | **Squid** | **Hooks min. 9 mm wide** | **Squid (*Loligo vulgaris*)** | **All year** |
| Trolling (*Fluixa*) | Hand line | Pelagic fishes | Hooks min. 9 mm wide | Mediterranean bonito (*Sarda sarda*), great amberjack (*Seriola dumerili*) | All year |
| Bottom hand line (*Volantí*) | Hand line | Demersal fishes | Hooks min. 9 mm wide | Red porgy (*Pagrus pagrus*), red scorpionfish | All year |
| | | | | (*Scorpaena scrofa*), comber (*Serranus cabrilla*), razorfish (*Xyrichtys novacula*) | |
| *Solta* trap net (*Solta*) | Fishing trap | Coastal fishes | 80 mm/ max. 300 m | Mediterranean bonito (*Sarda sarda*), great amberjack (*Seriola dumerili*) | All year |
| *Moruna* trap net (*Moruna*) | Fishing trap | Coastal fishes | 50 mm/ max. 500 m | Great amberjack (*Seriola dumerili*) | Spring-summer |
| *Almadraba* trap net (*Almadraba*) | Fishing trap | Coastal fishes | 200 mm/ max. 500 m | Great amberjack (*Seriola dumerili*) | Winter |
| Trawl (*Arrossegament*) | Bottom trawl | Benthic assemblages | 40 mm square mesh/ - | Red shrimp (*Aristeus antennatus*), Norway lobster (*Nephrops norvegicus*), hake (*Merluccius merluccius*), red mullet (*Mullus surmuletus*) | All year |

**Table 1**

Main characteristics of the 12 métiers initially identified. Expert fishers were asked to label a sample of daily boat records with one or more of these 12 métiers. The 7 métiers finally selected were denoted in bold. Note that an additional 13$^{th}$ category was considered for trawling because a few daily boat records from trawlers were erroneously included within the SSF data base.

| | Trawlers | Small-scale | Seiners |
|---|---|---|---|
| Trawlers | **218** | 0 | 0 |
| Small-scale | 0 | **176** | 0 |
| Seiners | 0 | 3 | **195** |

**Table 2**

Cross-validated confusion matrix for the classification algorithm intended to filter out trawlers and seiners. Successful predictions are at the main diagonal (in bold).

| | Cuttlefish/Fish | | Transparent goby | | Dolphinfish | | Squid jigging | | Spiny lobster | | Red mullet | | Longline | |
|---|---|---|---|---|---|---|---|---|---|---|---|---|---|---|
| | NO | YES | NO | YES | NO | YES | NO | YES | NO | YES | NO | YES | NO | YES |
| NO | **683** | 0 | **1045** | 0 | **1000** | 0 | **1044** | 0 | **867** | 0 | **1017** | 0 | **857** | 0 |
| YES | 5 | **412** | 0 | **55** | 0 | **100** | 0 | **56** | 2 | **231** | 1 | **82** | 0 | **412** |

**Table 3**

Cross-validated confusion matrix for the classification algorithm intended to predict métier from the daily boat record of landings. Note that in that case a binary classification was completed for each one of the métiers considered. Successful predictions are at the main diagonal (in bold).

| | TOTAL | Cuttlefish/Fish | Transparent goby | Dolphinfish | Squid jigging | Spiny lobster | Red mullet | Longline |
|---|---|---|---|---|---|---|---|---|
| Effort (fishing trips) | 14,652 | 4,964 (33.9% ) | 657 (4.5%) | 1,208 (8.2%) | 915 (6.2%) | 3,204 (21.9%) | 1,352 (9.2%) | 2,350 (16.0%) |
| Landings (kg) | 422,839 | 122,991 (29.1%) | 15,874 (3.8%) | 109,445 (25.9%) | 12,088 (2.9%) | 58,360 (13.8%) | 32,861 (7.8%) | 71,217 (16.8%) |
| Gross revenues (€) | 3,893,379 | 959,577 (24.6%) | 274,498 (7.1%) | 504,539 (13.0%) | 202,950 (5.2%) | 999,674 (25.7%) | 233,170 (6.0%) | 718,968 (18.5%) |

.

**Table 4**

Métier-specific average annual estimates of effort (fishing trips), landings (kg) and gross revenues (euros) of the small scale fleet from Mallorca between 2004 and 2015.

| MÉTIERS | | | | | | |
|---|---|---|---|---|---|---|
| Cuttlefish/Fish | Transparent goby | Dolphin fish | Squid jigging | Spiny lobster | Red mullet | Longline |
| SIPIA PT<br>*Sepia officinalis* | JONQUILLO<br>*Aphia minuta* | LLAMPUGA<br>*Coryphaena hippurus* | CALAMAR POT. PT<br>*Loligo vulgaris* | LLAGOSTA ROJA<br>*Palinurus elephas* | MOLL VERMELL GR<br>*Mullus surmuletus* | RATJADA<br>*Raja clavata* |
| VARIAT*<br>Mixed fish | JONQ./CABOTI<br>Mixed *A.minuta* and *Pseudaphya ferreri* | PAMPOL<br>*Naucrates ductor* | CALAMAR POT. GR<br>*Loligo vulgaris* | RATJADA<br>*Raja clavata* | MOLL VERMELL PT<br>*Mullus surmuletus* | DENTOL GR<br>*Dentex dentex* |
| SIPIA GR<br>*Sepia officinalis* | | VERDEROL<br>*Seriola dumerili* juvenile | | CAP ROIG MITJA<br>*Scorpaena scrofa* | VARIAT*<br>Mixed fish | PAGUERA PT<br>*Pagrus pagrus* |
| MORRALLA GR*<br>Mixed fish | | | | RAP MITJA<br>*Lophius budegassa* | MORRALLA GR.*<br>Mixed fish | DENTOL PT<br>*Dentex dentex* |
| POP GR<br>*Octopus vulgaris* | | | | CAP ROIG GR<br>*Scorpaena scrofa* | MORRALLA PT.*<br>Mixed fish | MORRALLA GR*<br>Mixed fish |

| | | | |
|---|---|---|---|
| ESCORPORA GR<br>*Scorpaena porcus* | CAP ROIG PT<br>*Scorpaena scrofa* | ESPARRALL<br>*Diplodus annularis* | CONGRE<br>*Conger conger* |
| CAP ROIG PT<br>*Scorpaena scrofa* | RATA<br>*Uranoscopus scaber* | POP MITJA<br>*Octopus vulgaris* | GATO<br>*Scyliorhinus canicula* |
| ESCORPORA PT<br>*Scorpaena porcus* | RAP PT<br>*Lophius budegassa* | RATA<br>*Uranoscopus scaber* | SIRVIOLA GR<br>*Seriola dumerili* |
| POP MITJA<br>*Octopus vulgaris* | RAP GR<br>*Lophius budegassa* | PAGELL PT<br>*Pagellus spp* | CANTERA GR<br>*Spondyliosoma cantharus* |
| RATA<br>*Uranoscopus scaber* | SIRVIOLA PT<br>*Seriola dumerili* | VAQUES/ VACAS<br>*Serranus scriba* | MUSSOLA<br>*Mustelus mustelus* |
| TORD<br>*Symphodus* spp. | GALL S.PEDRO MITJA<br>*Zeus faber* | | ORADA<br>*Sparus aurata* |
| PALOMIDA<br>*Lichia amia* | MOLLERA PT<br>*Phycis* spp. | | PAGUERA GR<br>*Pagrus pagrus* |

| | | |
|---|---|---|
| LLISSA<br><br>*Mugil spp* | PAGELL PT<br><br>*Pagellus* spp | ANFOS PT<br><br>*Epinephelus* spp. |
| ESCORBALL<br><br>*Sciaena umbra* | FERRASSA<br><br>*Dasyatis pastinaca* | MORENA<br><br>*Muraena helena* |
| SALPA<br><br>*Salpa salpa* | LLAGOSTA ROJA GR<br><br>*Palinurus elephas* | ARANYA CAP NEGRE<br><br>*Trachinus radiatus* |
| CARACOLA/ CORNET<br><br>*Trunculariopsis trunculus* | GALL S.PEDRO GR<br><br>*Zeus faber* | SARD MITJA<br><br>*Diplodus sargus* |
| GALL S.PEDRO PT<br><br>*Zeus faber* | LLAGOSTA ROJA MITJA<br><br>*Palinurus elephas* | SARD GR<br><br>*Diplodus sargus* |
| VARIADA<br><br>*Diplodus vulgaris* | CIGALA<br><br>*Scyllarides latus* | ANFOS GR<br><br>*Epinephelus* spp. |
| BURRO/ASE *Dactylopterus volitans* | PELAIA PT *Solea spp.* | SERRA *Serranus cabrilla* |
| POP PT *Octopus vulgaris* | | ESPET GR *Sphyraena viridis* |
| SURE *Balistes carolinensis* | | MOLLERA GR *Trisopterus minutus* |
| | | CANTERA PT *Spondilosoma cantarus* |

**Table 5**

List of the commercial categories that significantly contributed to define the seven métiers (i.e., results of the SIMPER analysis). First sale commercial categories (rows) for each métier (columns) for the small-scale fisheries from Mallorca (see Table 1 for the definition of métiers). The actual label (i.e., the label selected by the auctioneer from a closed list) is provided followed by the size when that determines the category. The species corresponding to each category is detailed. *GR*: big size; *MITJA*: medium size; *PT*: small size; * denotes the three commercial categories that contain several species.

Figure captions

**Fig. 1.** Analytical approach proposed for predicting the métier of each fishing trip from Mallorca Island's small scale fleet. The high numbers are those detailed in Section 2.1.

**Fig. 2.** Results of the multivariate analysis (CCA) aiming to check between-expert coherence. According to the expert-by-expert sequential elimination protocol (Step 8 in Section 2.1), the labeling of expert 6 did not match with the one of the other experts, hence it was deleted in the first loop. The four experts finally deleted are denoted in bold.

**Fig. 3.** Temporal trends for the effort (a), landings (b), gross revenues (c), landings per fishing trip (CPUE) (d) and gross revenues per fishing trip (e) for Mallorca's small scale fleet. Fishing trips have been polled by month.

**Fig. 4.** Fishing trips temporal trend distributed by métier. Separate panels denote effort (a), landings (b), and gross revenues (c). Fishing trips have been polled by month.

**Fig. 5.** Seasonal temporal trends (fishing trips of the same month are polled across the timeline considered in this study).

**Fig. 6.** Plot showing the optimal number of groups (i.e. métiers) in which the fishing trips are clustered. The vertical axis represents the average ratio between the similarities of a fishing trip with the centroid of its cluster, in relation to the similarity to other clusters. The horizontal axis represents the number of cluster considered. It is expected that the curve would peak at the optimal number of groups (2 métiers).

**Fig. 7.** General workflow proposed for predicting all the fishing trips métiers for other fisheries. According with Section 4.1, expert fishers' classification is preferred over the alternative pathways (denoted by dashed lines).

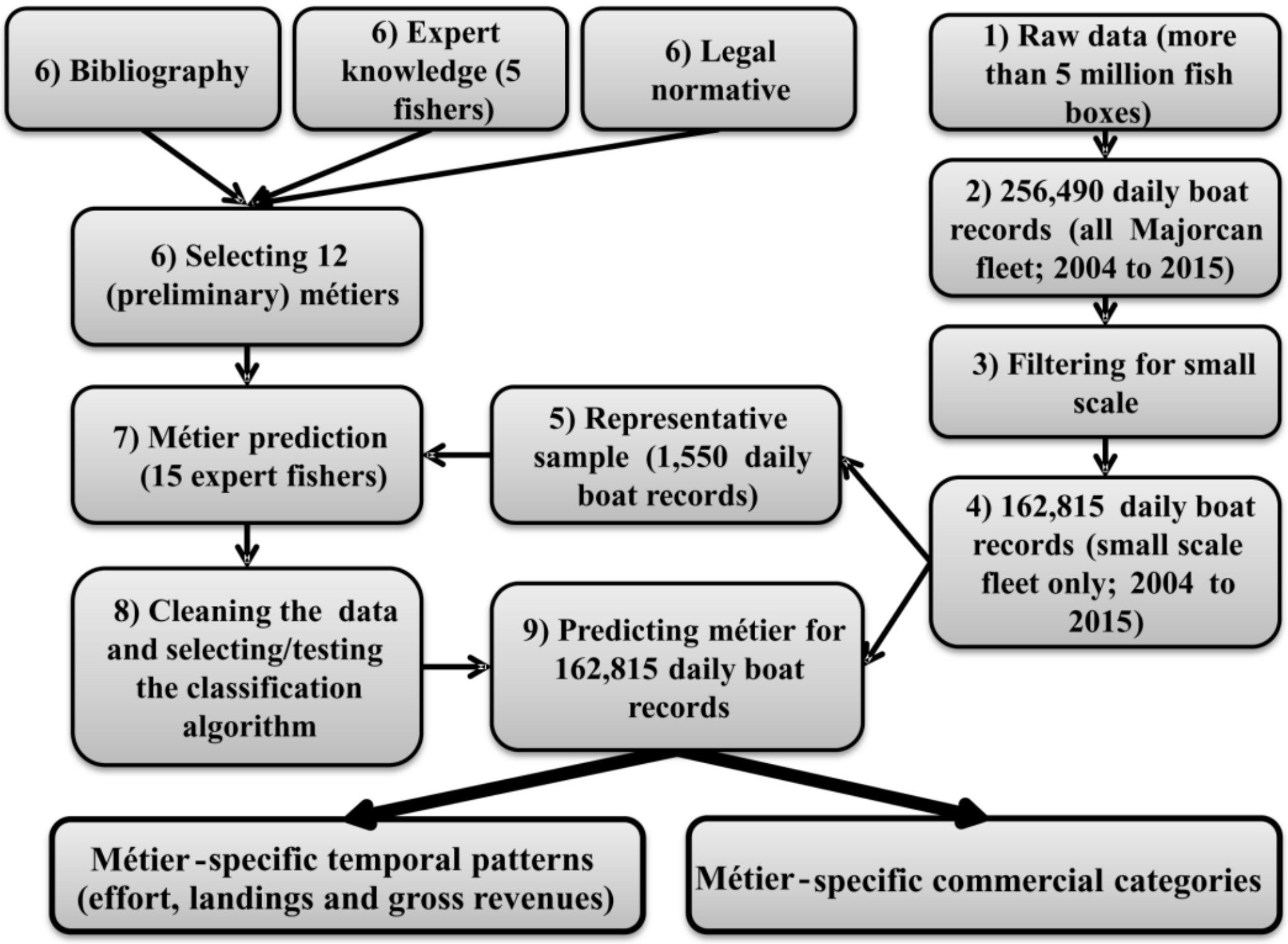
6) Bibliography
6) Expert knowledge (5 fishers)
6) Legal normative
1) Raw data (more than 5 million fish boxes)
2) 256,490 daily boat records (all Majorcan fleet; 2004 to 2015)
6) Selecting 12 (preliminary) métiers
3) Filtering for small scale
7) Métier prediction (15 expert fishers)
5) Representative sample (1,550 daily boat records)
4) 162,815 daily boat records (small scale fleet only; 2004 to 2015)
8) Cleaning the data and selecting/testing the classification algorithm
9) Predicting métier for 162,815 daily boat records
Métier-specific temporal patterns (effort, landings and gross revenues)
Métier-specific commercial categories

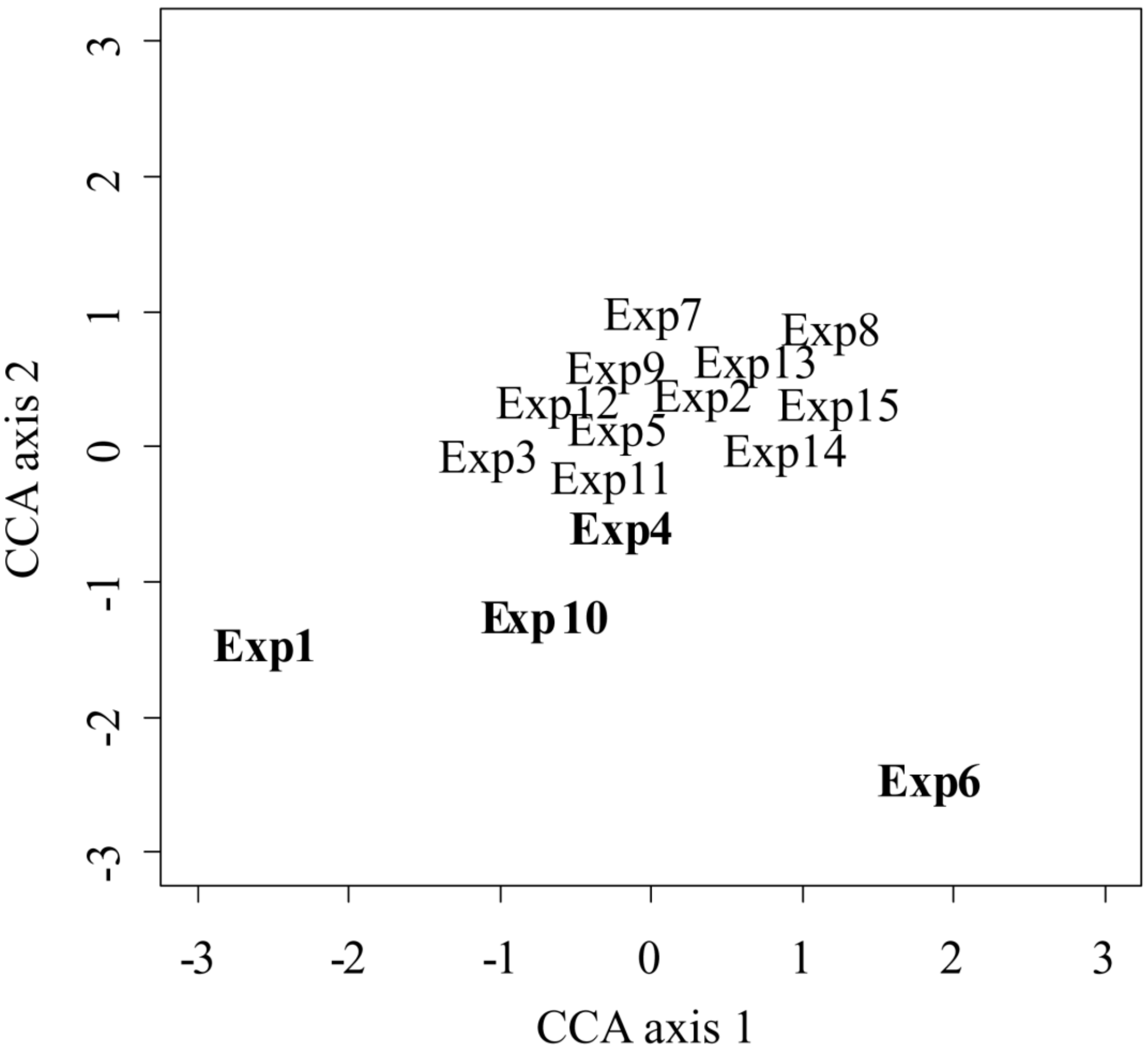

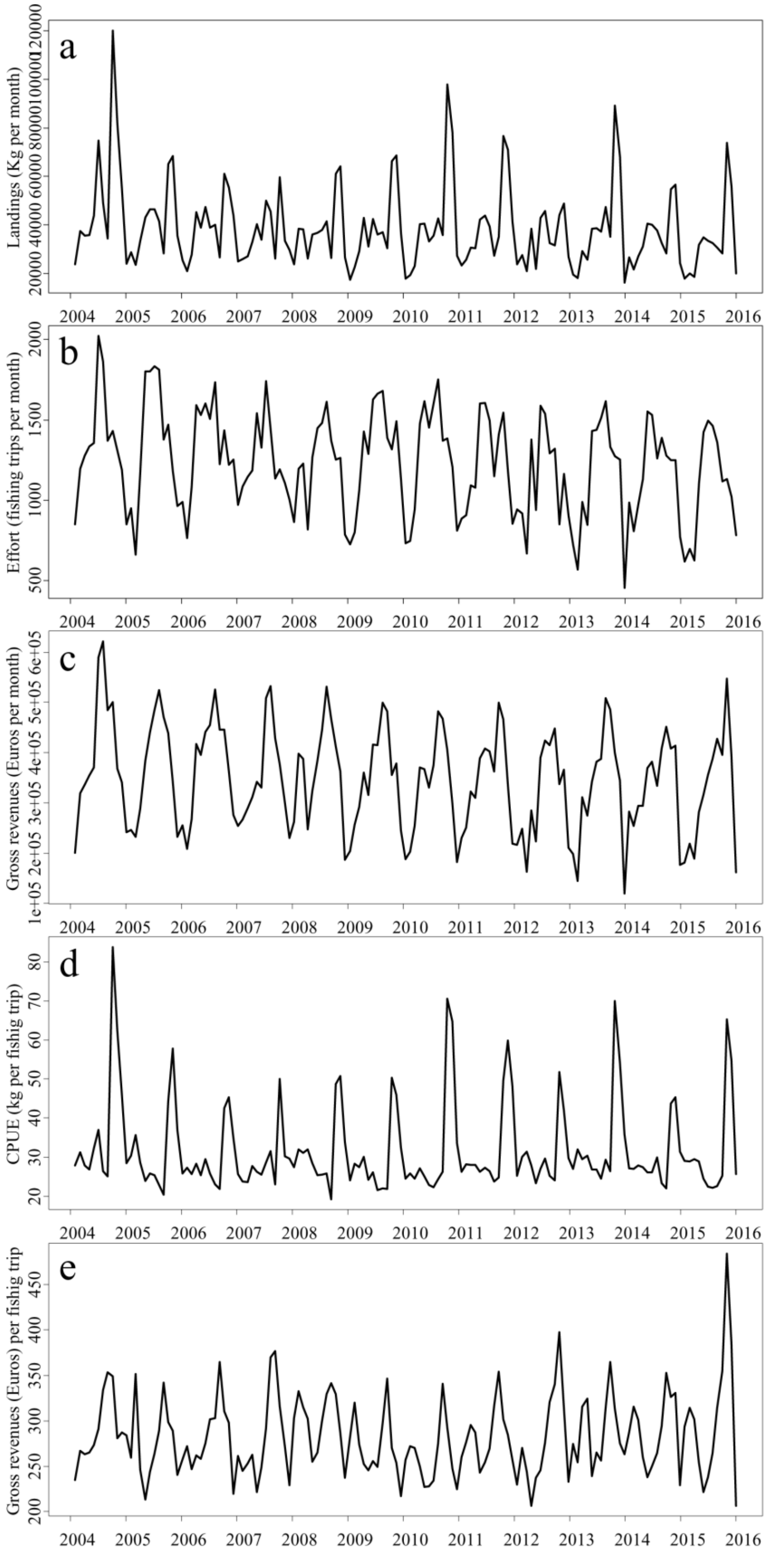

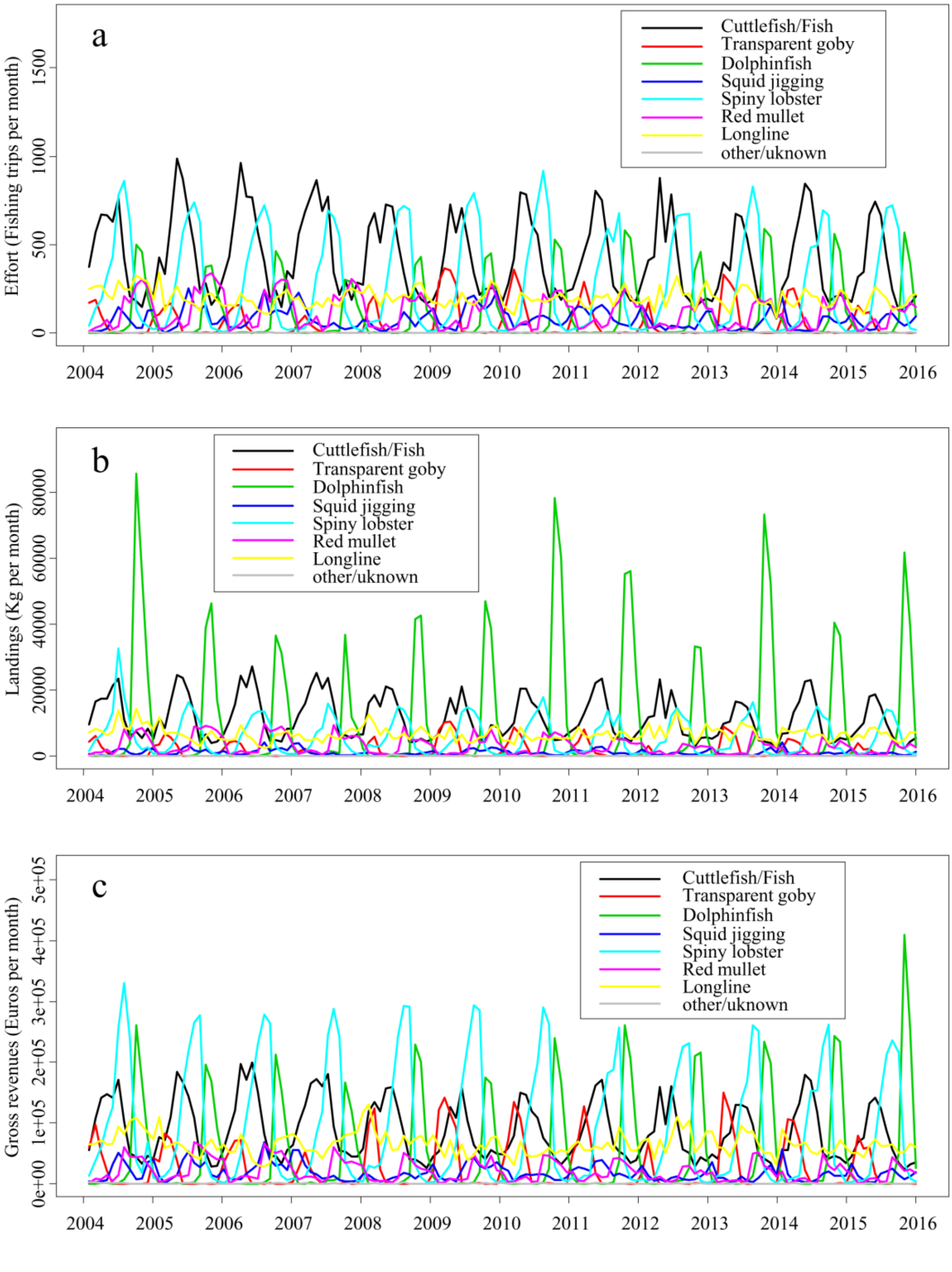

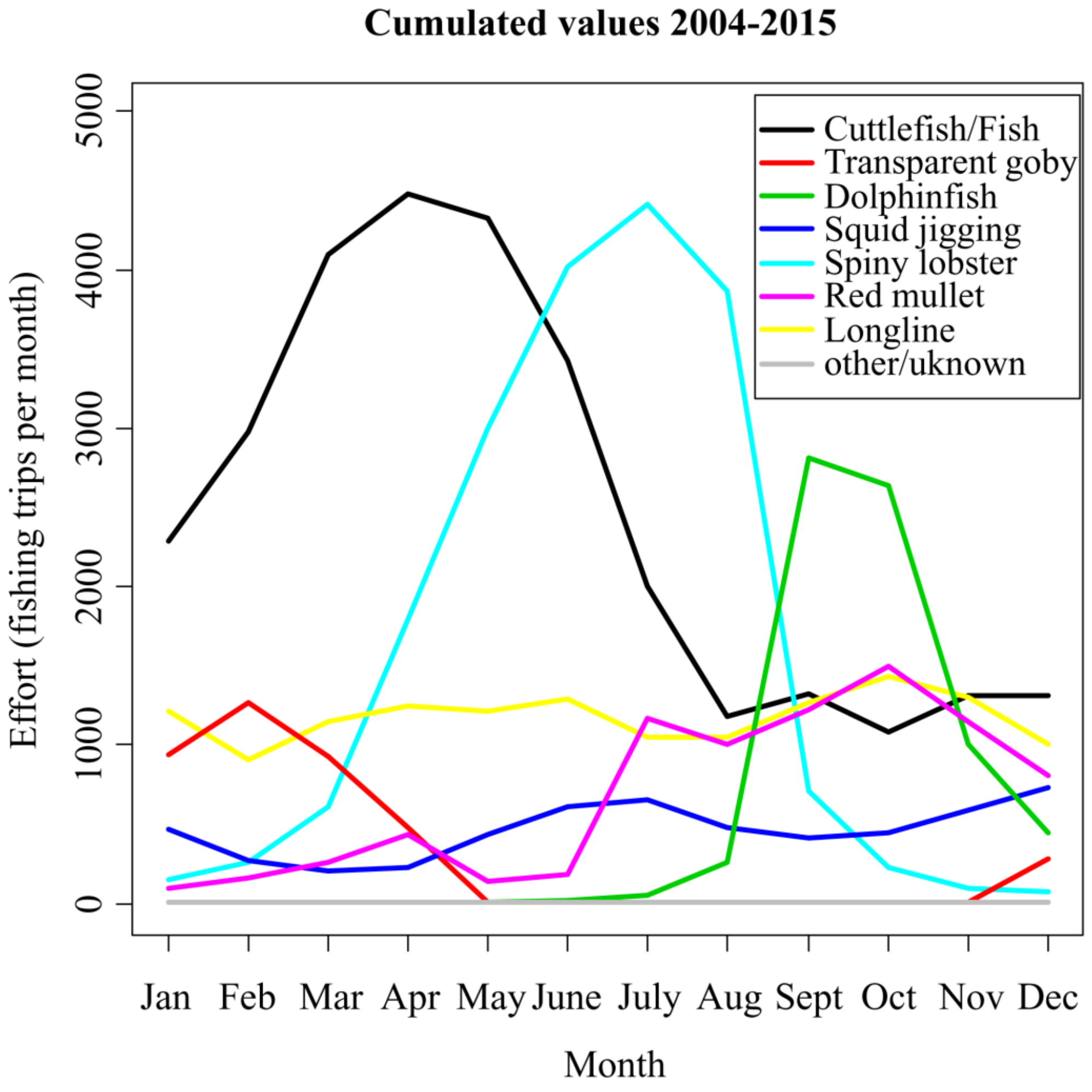

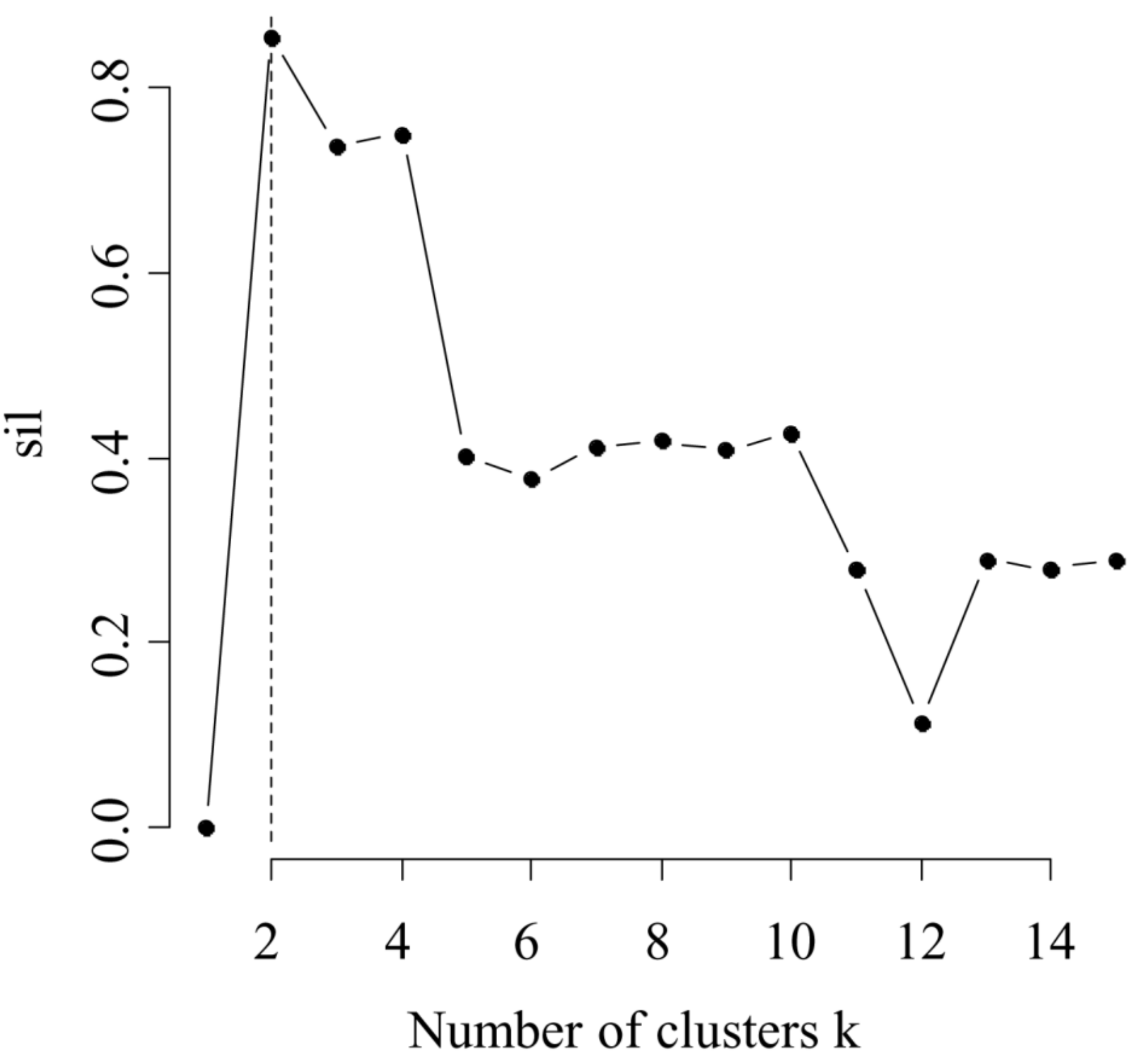

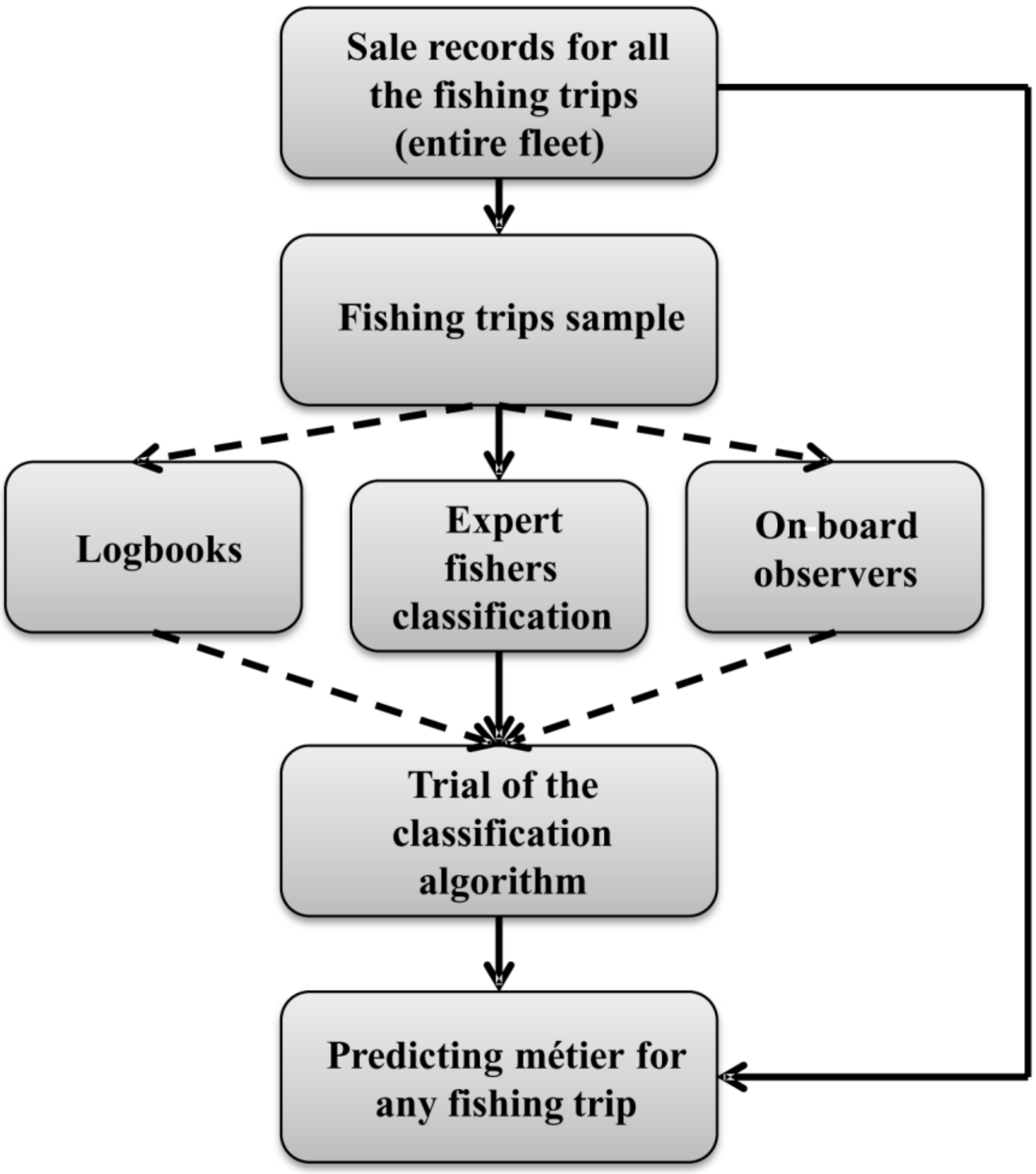
Sale records for all the fishing trips (entire fleet)
Fishing trips sample
Logbooks
Expert fishers classification
On board observers
Trial of the classification algorithm
Predicting métier for any fishing trip